\documentclass[aps,prd,twocolumn,superscriptaddress,amsfont,graphicx,nofootinbib,preprintnumbers]{revtex4}%
\usepackage{color,graphicx,epsfig}
\usepackage{ifpdf}
\usepackage{amsmath}
\usepackage{bm}
\usepackage{color}
\usepackage[english]{babel}
\usepackage{graphicx}%
\usepackage{amsfonts}%
\usepackage{amssymb}
\usepackage{braket}
\usepackage{hyperref}

\definecolor{nicered}{rgb}{0.7,0.1,0.1}
\definecolor{nicegreen}{rgb}{0.1,0.5,0.1}
\hypersetup{colorlinks,citecolor= nicegreen,linkcolor= nicered}

\newcommand{\beq}{\begin{equation}}
\newcommand{\eeq}{\end{equation}}
\newcommand{\bea}{\begin{eqnarray}}
\newcommand{\eea}{\end{eqnarray}}

\definecolor{Red}{rgb}{1.,0.,0.}

\newcommand{\tot}{\mathrm{tot}}
\newcommand{\high}{\mathrm{high}}
\newcommand{\low}{\mathrm{low}}

\begin{document}

\def\LjubljanaFMF{Faculty of Mathematics and Physics, University of Ljubljana,
 Jadranska 19, 1000 Ljubljana, Slovenia }
\def\Cincy{Department of Physics, University of Cincinnati, Cincinnati, Ohio 45221,USA}
\def\LjubljanaIJS{Josef Stefan Institute, Jamova 39, 1000 Ljubljana, Slovenia}
\def\Weizmann{Department of Particle Physics and Astrophysics,
Weizmann Institute of Science, Rehovot 76100, Israel}
\def\CERN{CERN, Theoretical Physics, CH-1211 Geneva 23, Switzerland}


\title{Forward Tevatron Tops and Backward LHC Tops with Associates}

\author{Jure Drobnak} 
\email[Electronic address:]{jure.drobnak@ijs.si} 
\affiliation{\LjubljanaIJS}

\author{Alexander L. Kagan}
\email[Electronic address:]{kaganalexander@gmail.com}
\affiliation{\Cincy}
\affiliation{\CERN}

\author{Jernej F.\ Kamenik} 
\email[Electronic address:]{jernej.kamenik@ijs.si} 
\affiliation{\LjubljanaIJS}
\affiliation{\LjubljanaFMF}

\author{Gilad Perez}
\email[Electronic address:]{gilad.perez@cern.ch}
\affiliation{\Weizmann}
\affiliation{\CERN}

\author{Jure Zupan} 
\email[Electronic address:]{jure.zupan@cern.ch} 
\affiliation{\Cincy}

\date{\today}
\begin{abstract}
The anomalous forward-backward asymmetry ($A_{FB}$) in $t\bar t$ production at the Tevatron is in apparent contradiction with the $t\bar t$ charge asymmetry ($A_C$) measurements at the LHC, which agree well with the standard model predictions. 
We argue that associated production of a state with $[u\bar t]$ flavor quantum numbers
can lead to a sizable negative contribution to $A_C$. 
Exchange of such a resonance in the $t-$channel would lead to positive contributions to  $A_{FB}$  and $A_C$, as has been extensively discussed in the literature.
Given the additional negative $A_C$ contribution, both the Tevatron and LHC data can be naturally accommodated within this framework.  
A simple realization of this setup 
is the well known example of a $Z'$ with flavor off-diagonal up-top couplings.
We provide a detailed study of this model, demonstrating that it indeed reproduces the $t\bar t$ asymmetry data, and  is compatible with other constraints, {\it e.g.} $t+{\rm jet}$ resonance searches, $t\bar t$ inclusive jet multiplicities, and atomic parity violation.
\end{abstract}

\maketitle

\section{Introduction} 
The inclusive forward-backward asymmetry ($A_{FB}$) in the $t\bar t$ rest frame has been measured by both the CDF~\cite{AFBCDF,AFBCDF1} and D\O~\cite{AFBD0} collaborations and found to be significantly larger than the standard model (SM) prediction. The na\"ive average in semileptonic $t\bar t $ events, 
adding the uncertainties in quadrature, is
\beq
A_{FB} = 0.174\pm 0.038\,.
\eeq
The NLO QCD predictions, including leading electroweak (EW) contributions~\cite{Hollick,Kuhn,Manohar,Bernreuther:2012sx},
are~\cite{Bernreuther:2012sx}
\beq A^{\rm SM}_{FB} = 0.088 \pm 0.006\,, \label{AFBsm} \eeq
when using the LO total cross section in the denominator of the asymmetry. The error is due to scale variation with the central value evaluated at 
$\mu = m_t$ and the low and high values evaluated at $\mu = m_t /2$ and $ 2 m_t$, respectively.
Both CDF and D\O \, have also measured the asymmetry in bins of the $t\bar t$ invariant mass and $t\bar t$ rapidity difference. Only CDF~\cite{AFBCDF,AFBCDF1}, however, unfolds to the partonic (``truth") level obtaining
\begin{subequations}
\begin{align}
A_{FB}^{\rm low}\equiv A_{FB}(m_{t\bar t}<450~{\rm GeV}) &= 0.078 \pm 0.054\,,\\
A_{FB}^{\rm high}\equiv A_{FB}(m_{t\bar t}>450~{\rm GeV}) &= 0.296 \pm 0.067\,,
\end{align}
\end{subequations}
to be compared with the SM (NLO QCD and EW) predictions~\cite{Bernreuther:2012sx,Hollick,Kuhn} 
\beq (A^{\rm low}_{FB})^{\rm SM} = 0.062^{+ 0.004}_{-0.003},~~~(A^{\rm high}_{FB})^{\rm SM}= 0.129^{+0.008}_{-0.006}\,, \label{AFBsmhighlow}\eeq
again using the LO cross sections in the asymmetry denominators, with the errors due to scale variation as above.
At the top reconstruction level, D\O \, does not observe a significant rise in the asymmetry vs. the $t\bar t$ invariant mass, however an increase vs. the $t\bar t$ rapidity difference is observed 
~\cite{AFBD0}.

A related observable at the LHC is the charge asymmetry in $t\bar t$ production, $A_C$. In contrast to the forward-backward asymmetries, the measurements of $A_C$ agree with the SM expectations. For example, the average of the inclusive ATLAS~\cite{ACATLAS} and CMS~\cite{ACCMS} measurements at $\sqrt{s}=7$ TeV, 
\beq\label{ACinclusive}
A_C = (1.15 \pm 1.25)\cdot 10^{-2} \,,
\eeq
agrees within errors with the 
SM prediction~\cite{Bernreuther:2012sx,Kuhn}
\beq  A_C^{\rm SM} = (1.23\pm 0.05)\cdot 10^{-2}\,. \label{ACsm}\eeq
ATLAS collaboration also presented the first measurements of $A_C$ in bins of $m_{t\bar t}$~\cite{ACATLAS} 
\begin{subequations}
\begin{align}
A_C^{\rm low}\equiv A_C(m_{t\bar t}<450~{\rm GeV}) &= -0.053 \pm 0.088\,,\\
A_C^{\rm high}\equiv A_C(m_{t\bar t}>450~{\rm GeV}) &= -0.008 \pm 0.047\,,
\end{align}
\end{subequations}
which are consistent with the SM predictions~\cite{Kuhn}, $A_C^{\rm low} = 0.015\pm 0.001$ and $A_C^{\rm high}= 0.026\pm 0.001$
(all LHC asymmetry predictions correspond to LO cross sections in the denominators, and the errors are due to scale variation, as above).
Compatible results, but in three $m_{t\bar t}$ bins, were presented by the CMS collaboration~\cite{ACCMS}. 

The discrepancy between the SM predictions and the measured asymmetries at the Tevatron could be due to an unknown QCD effect,  or an unidentified experimental bias. Alternatively, it might be a hint of dynamics beyond the SM (for a review, see, e.g.,~\cite{Kamenik:2011wt}).  
However, new physics models are challenged by the requirement of achieving simultaneous agreement with the anomalously large $A_{FB}$ measurements and the SM-like $A_C$ measurements, as the
underlying candidate NP processes often yield correlated positive contributions to both 
\cite{AguilarSaavedra:2011hz,Cao:2011hr,Fajfer:2012si}.\footnote{However, in the case of a light $s$-channel axigluon, it has recently been shown that different couplings to the $u$ and $d$ quarks can lead to partial cancelations between the $u\bar u$ and $d\bar d$ contributions, thus sufficiently suppressing $A_C$, with only a small impact on $A_{FB}$, which can be sizable and positive
\cite{Drobnak:2012cz,AguilarSaavedra:2012va}.}

\begin{figure}[t]
\centering{
\includegraphics[scale=0.55]{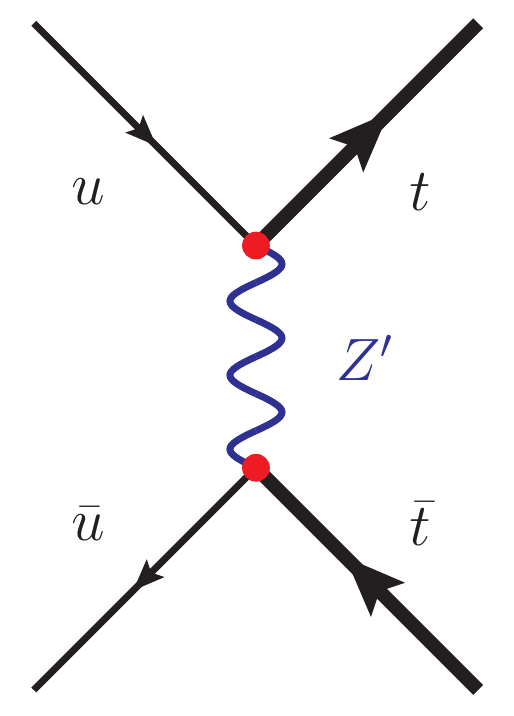} 
\includegraphics[scale=0.55]{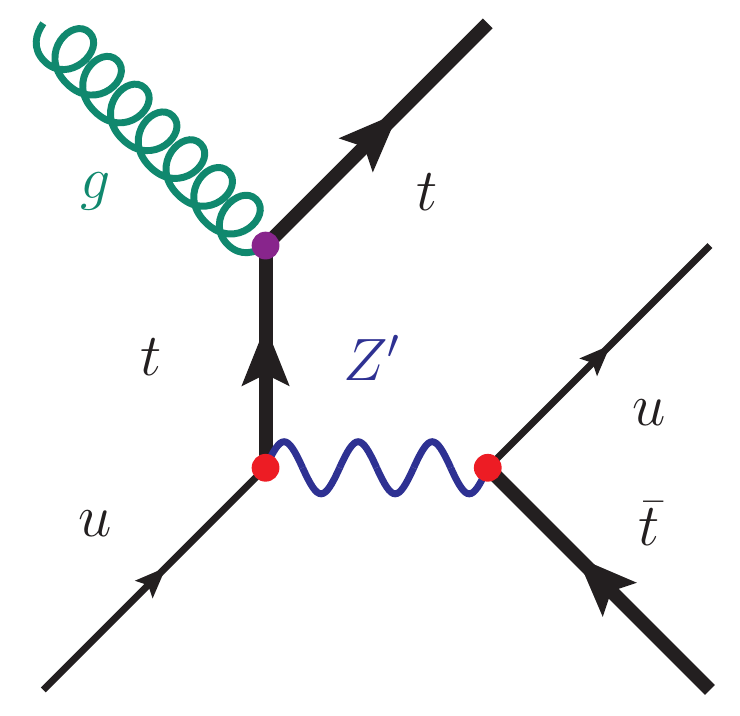} 
\includegraphics[scale=0.55]{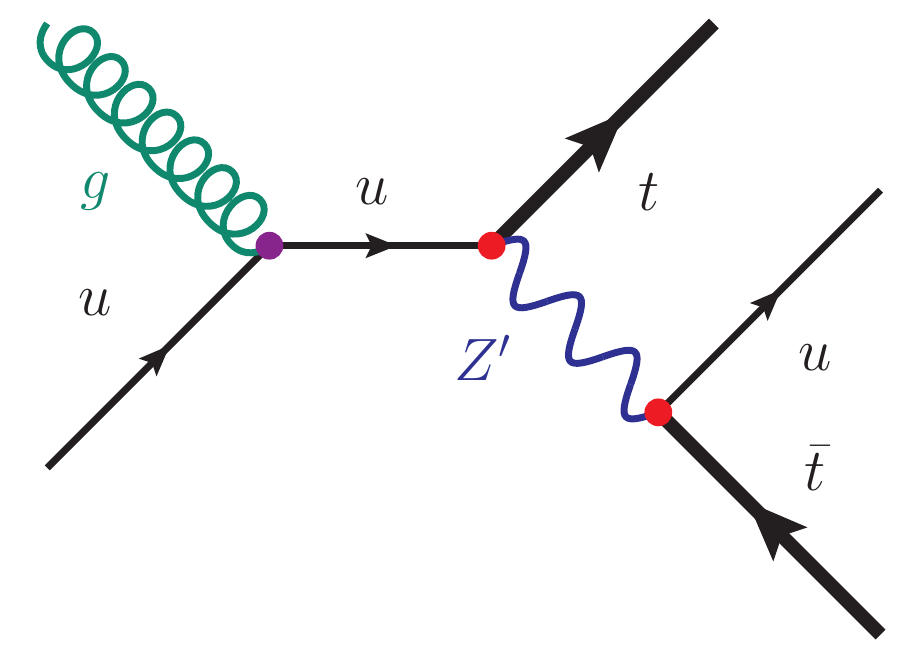} 
}
\caption{Feynman diagrams for (a) $t$-channel $Z'$ exchange contribution to $u\bar u \to t\bar t$, and
associated single $Z'$ production in the $t$ channel (b) and $s$ channel (c). 
\label{fig:feyn}}
\end{figure}

The main purpose of this paper is to point out that there is a class of models in which positive contributions to $A_{FB}$ and $A_C$ are correlated with an additional negative contribution to $A_C$. Therefore, the latter tends to be small, in agreement with current data.  
A state with $[\bar  t u]$ or $[\bar t d]$ flavor quantum numbers is required.  
The possible examples are the $[\bar t  u ]$ flavored $Z^\prime$  \cite{Jung:2009jz} (see also~\cite{Ko:2011di}) or $[\bar t d]$ $W^\prime$ \cite{Cheung:2009ch} vector mediators, 
and the  $[\bar t  u ]$ flavored $SU(2)_L$ doublet scalar mediator $\phi$ \cite{Blum:2011fa}.   
In this paper we focus on the $Z^\prime$ color- and weak-singlet, with a coupling to the right-handed up and top quarks.
Exchange of the $Z'$ in the $t$-channel, see Fig.~\ref{fig:feyn}a, would lead to an increase in $A_{FB}$ vs. $m_{t\bar t}$, and a positive contribution to $A_C$ in excess of the measurements  \cite{AguilarSaavedra:2011hz,Fajfer:2012si}, due to forward peaking.  However,
associated production of the $Z'$ with a top-quark would produce an additional negative contribution to $A_C$.
Presumably, this would also be the case for the $W^\prime $ and $\phi$.  However, we leave the question of the overall viability of these models for future studies.

\begin{figure}[t]
\centering{
\includegraphics[scale=0.6]{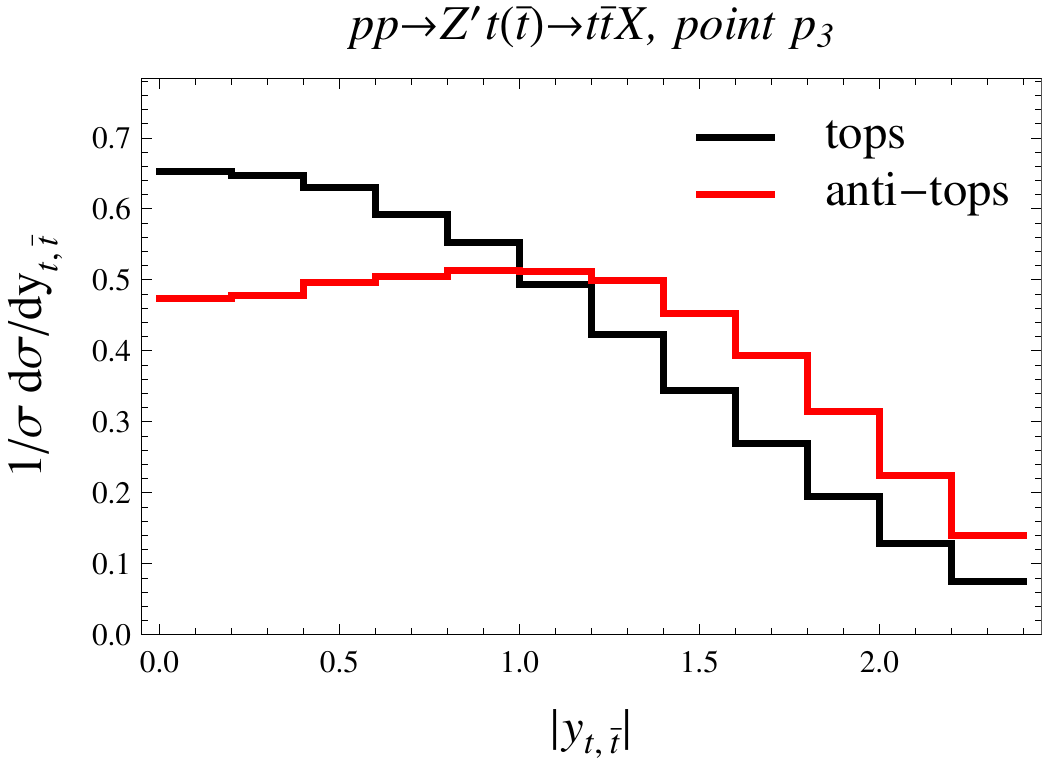} 
\includegraphics[scale=0.6]{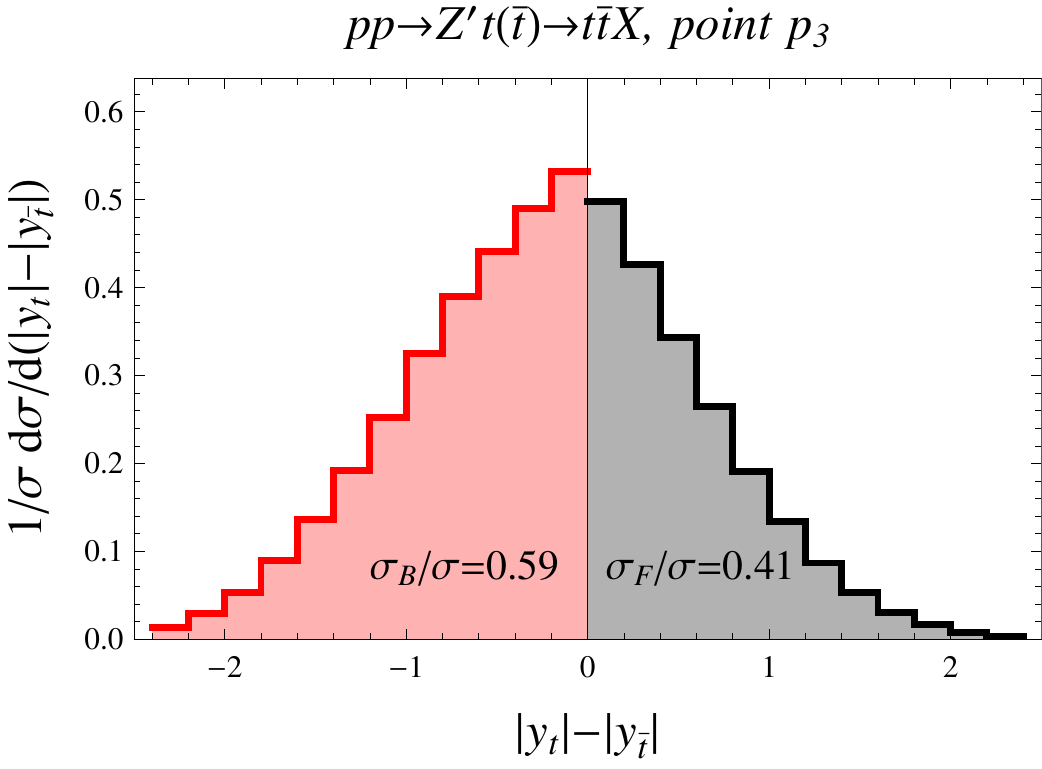}}
\caption{The relative differential cross sections 
(a) $1/\sigma ~d\sigma / d {|y_{t,\bar t}|}$ and (b)  
$1/\sigma ~d\sigma /  d {( |y_{t}| -    |y_{\bar t} | )}$ 
for $pp \to Z^\prime t (\bar t) \to t \bar t X $ at $\sqrt{s} = 7$ TeV, for benchmark point $p_3$ in Table~\ref{tab:BMP}.
The subscripts $B$ and $F$ refer to $|y_t | - |y _{\bar t} | <  0$ and $> 0 $, respectively.
\label{fig:rapidity}}
\end{figure}

The leading order $ ug \to Z't \to u\bar t t$ Feynman diagrams are shown in Figs.~\ref{fig:feyn}b and~\ref{fig:feyn}c. 
The effect we are interested in is due to the top quark exchange diagram in Fig.~\ref{fig:feyn}b.
The $Z'$ decay yields a $\bar t$ quark which tends to be boosted in the same direction as the incoming 
$u$ quark.  Taking into account the harder $u$ quark vs.  gluon PDF's in the proton, one concludes that on average 
the $\bar t$ is produced with a larger rapidity than the $t$, thus yielding a negative contribution to $A_C$.
The efficiency of this mechanism is illustrated in 
Fig.~\ref{fig:rapidity} for one of the $Z^\prime$ benchmark points ($p_3 $) listed in Table~\ref{tab:BMP}.
The $pp \to  (Z' t, Z^{\prime\dagger} \bar t)   \to \bar t t X$ differential cross section plots in Fig. \ref{fig:rapidity}a and Fig. \ref{fig:rapidity}b, respectively, exhibit the dominance of $\bar t$ ($t$) production at larger (smaller) rapidities and 
the dominance of $t\bar t$ production for $|y_t | -  |y_{\bar t} | <0$.  According to Fig. \ref{fig:rapidity}b, 
the charge asymmetry from associated vector mediator production alone can be large and negative, e.g., $A_C= - 18\%$ for the example shown.

At the LHC, the cross section for the CP conjugate process, $\bar u g \to Z^{\prime \dagger}  \bar t \to \bar u t\bar t$, is typically an order of magnitude smaller, due to the $\bar u $-quark PDF in the initial state.   This asymmetry has been studied previously in \cite{Knapen:2011hu}.  A corollary is that the negative contribution to $A_C$ does not depend on whether the $Z'$ is self conjugate or not.
However, bounds on same sign top production can rule out a self conjugate $Z^\prime$.
In fact, flavor symmetric realizations of the vector $t$-channel models \cite{Jung:2011zv,Grinstein:2011dz1,Grinstein:2011dz2} 
proposed to trivially evade  
the bounds on same sign top or single top production and FCNC's, e.g., $D-\bar D$ mixing, 
would contain a CP conjugate pair of $[\bar t  u]$ and $[t \bar u]$ flavored $Z'$'s.  
At the Tevatron, associated production of the vector mediators produces a negative contribution to $A_{FB}$.   However,  this effect is suppressed relative to the positive $A_{FB}$ contribution from Fig.~\ref{fig:feyn}a by the smaller gluon vs. $u $-quark PDF's inside the proton.

It is worth pointing out that in models for enhanced $A_{FB}$ in which a mediator is exchanged in the $u$-channel rather than the $t$-channel, e.g.,
a color triplet or sextet scalar $S$~\cite{Shu:2009xf,Grinstein:2011dz2,Dorsner:2009mq,Ligeti:2011vt,Grinstein:2011dz1}
with couplings $S\, t_R \,u_R$ (where the color indices have been suppressed), single mediator production via the analog of Fig.~\ref{fig:feyn}b would lead to a positive contribution to $A_C$~\cite{Hagiwara}.  This is easily understood, as in this case the mediator would decay to a top quark rather than an anti-top quark.

Associated $Z^\prime$ production
is subject to several constraints which need to be checked in order to establish the viability of our mechanism for 
reducing $A_C$. The most relevant ones are: {\em i)} the LHC $t\bar t$ cross section measurement, which constrains the product
$\sigma (pp \to Z' t) \times {\rm Br}(Z' \to t \bar u )$~\cite{KZassociatedprod};  
{\em ii)} the CDF, CMS and ATLAS collaborations $Z^\prime \to t+ {\rm jet}$ resonance searches~\cite{Aaltonen:2012qn,Chatrchyan:2012su,ATLAStj}, that yield direct bounds on the above product~\cite{Gresham:2011dg};
{\em iii)} the CMS measurement of the jet multiplicity distribution in semileptonic $t\bar t$ events, which is consistent with SM Monte Carlo studies~\cite{CMSjets}, thus potentially constraining NP models which produce extra jets~\cite{Sheltontalk}.

Constraints {\em ii)} and {\em iii)} become weaker as the $Z'$ mass approaches $m_t$ from above.  Because the $u$ quark becomes softer, the resonance searches lose sensitivity and the jet multiplicities become more SM like.  
Furthermore, a reduction of ${\rm Br}(Z' \to t \bar u )$
trivially reduces the impact of searches {\em i)}-{\em iii)}. 
In fact, we will see that the ranges $M_{Z'} \sim 200 - 300$ GeV 
and ${\rm Br}(Z' \to \bar t u) \sim 1/4 - 1/3$  are preferred, with the best overall fits occurring for $M_{Z'} $ near 220 GeV.  
A sufficient negative shift in $A_C$ can be obtained, after taking the branching ratio into account,
due to the sizable negative charge asymmetry intrinsic to associated $Z^\prime (\to \bar t u)$ production.

The favored range for ${\rm Br}(Z' \to \bar t u)$ raises an immediate question: what are the viable candidates for the missing dominant $Z'$ decay?
Some possibilities that come to mind are invisible decays, decays to light quark pairs, and decays to a pair of scalars
which subsequently decay to quark pairs, which could include a virtual top quark.\footnote{The last option 
could be realized in flavor symmetric models based on new strong interactions, in which the $Z'$'s are composite elements of a flavor nonet, analogous to the $K^{*\pm}$ of QCD, with dominant decays to pseudo Nambu-Goldstone bosons~\cite{stronginteractions}.}
The first option yields mono-top  events~\cite{Andrea:2011ws,Kamenik:2011nb}.
The CDF collaboration performed a search for the monotop signal and obtains a bound on the cross section of $0.5$\,pb based on $7.7\,{\rm fb}^{-1}$ of data~\cite{Aaltonen:2012ek} (for a $Z'$ mass between $0$ and $150$ GeV, while the bound would be weaker for a heavier $Z'$). This means that all of the benchmarks to be discussed below are allowed to decay predominantly into invisible states.  
The second and third options result in a $t+n~ {\rm jets}$ final state ($n\geq 2)$. Single top searches at the Tevatron~\cite{Abazov:2009ii,Abazov:2009pa,Aaltonen:2010jr,Aaltonen:2009jj} and the LHC~\cite{Chatrchyan:2011vp,Aad:2012ux}
have not been optimized for the higher number of high $p_T$ jets compared to the SM. Nevertheless they may pose an important constraint.
A detailed investigation of these issues is left 
for future work.

Finally, the $Z'$ model is subject to low energy constraints from atomic parity violation (APV) measurements~\cite{KZAPV}, as are the $W'$ and $\phi$ models. 
As it stands, the $Z'$ model only contains massive vectors with flavor violating couplings to quarks. Thus, a UV completion would be required in order to render it renormalizable,
and similarly for the $W'$ model.
As we shall see, at the level of an effective theory in which one only considers the effects of the 
$Z'$, the severity of these constraints significantly weaken (we also comment below on possible UV completion of our model).
A shift in the Cs weak nuclear charge $Q_W (^{133} {\rm Cs})$ due to the nuclear matrix element corrections recently discussed in~\cite{Dzuba:2012kx} 
also ameliorate the APV constraints. 
Thus, one can not claim any tension with the APV data in the preferred region of parameter space for the $Z'$ model.

The organization of the paper is as follows.  In Section~\ref{sec:2} we discuss our fits to the Tevatron and LHC $t \bar t $ production data, thus arriving at the 
preferred region of parameter space for the $Z'$ model.  We will demonstrate that the addition of associated 
$Z'$ production is crucial for obtaining a good fit, due to the LHC measurements of $A_C$.  This is illustrated in Fig.~\ref{fig:row1}, 
which is the main result of this paper.  We also check that the predicted Tevatron and LHC $m_{t\bar t}$
spectra agree with the measured ones.
In Section~\ref{sec:3} we discuss the impact of the $t j$ resonance searches and jet multiplicity measurements.
Atomic parity violation is discussed in Section~\ref{sec:4}.  In Section~\ref{Lepton} we
briefly discuss the implications of the $Z'$ model for the leptonic forward-backward and charge asymmetries, and conclude in Section\ref{sec:5}.


\section{Importance of associated $Z'$ production}
\label{sec:2}

The Lagrangian for the $Z'$ model is given by 
\beq   {\cal L}= g_{ut}  \, Z^\prime_\mu \,  \bar u_R \,\gamma^\mu \, t_R  \,+ \,{\rm h.c.} \,  + \,M_{Z^\prime}^2 \, Z^{\prime\,\dagger}_\mu \,Z^{\prime\, \mu}  .\label{eq:lagrangian} \eeq
As previously mentioned, the $Z^\prime $ would not be self-conjugate if it transformed non-trivially under
flavor symmetries introduced to suppress same sign top or single top production and FCNC's~\cite{Jung:2011zv,Grinstein:2011dz1,Grinstein:2011dz2}. However, a self-conjugate $Z'$ would not modify the results of our analyses for $t \bar t$ observables.
We employ a $\chi^2$ function to search for optimal ranges of the coupling $g_{ut}$, the vector mass $M_{Z^\prime}$, and ${\rm Br}(Z' \to t \bar u )$.
Six $t\bar t $ production observables enter our fit:  the total Tevatron and LHC cross sections, $\sigma^{\rm TEV}_{\rm tot}$ and 
$\sigma_{\rm tot}^{\rm LHC}$, 
the Tevatron inclusive forward-backward asymmetry $A_{FB}$, the ones for 
$m_{t\bar t }\ge  450$ GeV and $m_{t\bar t} <$ 450 GeV denoted as $A_{FB}^{\rm low}$ and $A_{FB}^{\rm high}$ respectively, 
and the LHC inclusive charge asymmetry $A_C$.  The experimental values are listed in the second column of Table~\ref{tab:expsmtheory}.  
Where more than one measurement is indicated we take the na\"ive average, with errors added in quadrature.

The NP contributions to the above observables are evaluated at LO, whereas
the SM contributions are taken at the highest order available.
The total cross sections are expressed as
\beq
\sigma_{\tot} =  \sigma^{t\bar t , Z^\prime }_{\tot} +  \sigma^{Z^\prime t}_{\tot} \times \mathrm{Br} (Z^\prime \to \bar t u )+ \sigma^{\rm SM}_{\tot} \,,
\label{sigmatot} 
\eeq
where $\sigma^{t\bar t , Z^\prime }$ denotes contributions from t-channel $Z^\prime$ exchange, including 
interference with the SM amplitude at LO, and 
$\sigma^{Z^\prime t} $ denotes the sum of the cross sections for $Z^\prime t$ and $Z^{\prime\,\dag} \,\bar t$ production.
The inclusive $t\bar t $ asymmetries are expressed as 
\beq 
A =\frac{ \Delta \sigma^{t\bar t , Z^\prime } + \Delta \sigma^{Z^\prime t }  + \Delta  \sigma^{\rm SM}  }{\sigma^{t\bar t , Z^\prime }_{\tot} +  \sigma^{Z^\prime t}_{\tot} \times \mathrm{Br} (Z^\prime \to \bar t u )+ \sigma^{\rm SM,\,LO}_{\tot} }\,.~~~~\label{AFB}
\eeq
For $A_{FB}$, 
\begin{eqnarray}
 \Delta  \sigma^{\rm SM}  &=&
 \sigma^{\rm SM}_F - \sigma^{\rm SM}_B\,,~~~\Delta \sigma^{t\bar t , Z^\prime }  = \sigma^{t\bar t , Z^\prime}_F - \sigma^{t\bar t , Z^\prime}_B,\label{dsigma1}\\
 \Delta \sigma^{Z^\prime t } &= &  \sigma_F \left(p \bar p \to Z^\prime t / Z^{\prime \,\dag}\,\bar t  \to t \,\bar t \,X \right) - \left(\sigma_F\to \sigma_B\right) ,~~~~~\label{dsigma2}
\end{eqnarray}
where the $F$ and $B$ subscripts denote events in which the top quark is emitted in the forward and backward hemispheres, respectively, with respect to the proton beam direction.  Note that $ \Delta \sigma^{Z^\prime t }$ scales as $\mathrm{Br} (Z^\prime \to \bar t u )$. The expressions for $A_{FB}^{\rm low}$ and $A_{FB}^{\rm high}$ correspond to the restrictions 
$m_{t\bar t} < 450 {\rm\,\,GeV}$ and $\ge 450 $ GeV, respectively, for the cross sections entering (\ref{sigmatot})--(\ref{dsigma2}).
For $A_C$, 
substitute $p\bar p \to pp$,
$F \to |y_t | - |y_{\bar t} | > 0$ and $B\to |y_t | - |y_{\bar t} | < 0$ in (\ref{dsigma1}), (\ref{dsigma2}), where $y_t$ and $y_{\bar t}$ are the $t$ and $\bar t$ rapidities.

\begin{table}
\begin{center}
\begin{tabular}{llll}\hline\hline
 \multicolumn{2}{c}{~~~~~~~~~~Experiment~~~~~~~~~~} & \multicolumn{2}{c}{~~~~~SM Theory~~~~} \\ \hline
$\sigma_{\tot}^{\mathrm{TEV}}$[pb]&$7.50\pm 0.48 $~\cite{CDF1} & $\sigma_{\tot}^{\mathrm{TEV}}$[pb] & $7.067 \pm 0.26$~\cite{Baernreuther:2012ws}\\
$A_{FB} $ [\%]& $17.4 \pm 3.8$~\cite{AFBCDF1,AFBD0} & $\Delta \sigma^{\rm SM }$[pb]  & $0.71,0.48,0.33,0.25$ \\
$A_{FB}^{\high}$[\%]& $29.6\pm 6.7$~\cite{AFBCDF1} &$\Delta \sigma^{\rm SM }$[pb]  &$0.46,0.27,0.18,0.13 $ \\
$A_{FB}^{\low}$[\%]&$7.8\pm 5.4$~\cite{AFBCDF1}&$\Delta \sigma^{\rm SM }$[pb]  & $0.27,0.22,0.15,0.11$ \\ \hline
$\sigma_{\tot}^{\mathrm{LHC}}$[pb]& $172.8\pm 8.2$~\cite{muller_ICHEP} &$\sigma_{\tot}^{\mathrm{LHC}}$[pb]& $162.6\pm 17.1$~\cite{Beneke:2011mq} \\
$A_C$[\%]& $1.15 \pm 1.25$\cite{ACATLAS,ACCMS} &$\Delta \sigma^{\rm SM }$[pb]  &$2.51,1.60,1.16,0.96$ \\\hline\hline
\end{tabular}
\caption{Experimental and SM predicted values for $t\bar{t}$ cross-sections and asymmetries at the Tevatron and LHC. Multiple entries correspond to $\Delta \sigma^{\rm SM}$ cross section differences evaluated at $\mu=m_t/2,m_t,2m_t,4m_t$, see text.  
} 
\label{tab:expsmtheory}
\end{center}
\end{table}

The $t\bar t$ cross section for the $i$'th $m_{t \bar t}$ bin
is written as
\beq
\sigma_{i} =  \left[\sigma^{t\bar t , Z^\prime }_{i}  +  \sigma_i \left(p \bar p \to Z^\prime t / Z^{\prime \,\dag}\,\bar t  \to t \,\bar t \,X \right) \right]\epsilon_{ i} + \sigma^{\rm SM}_{i}  \,,
\label{sigmai} 
\eeq
where the $\epsilon_{i}$ are acceptance corrections, relative to the SM ones, which account for a significant decrease in the CDF detector's sensitivity to top quark production in the forward region.  As the forward region features prominently at large invariant masses in the $Z'$ model, it is crucial to take these corrections into account when comparing the predicted $m_{t\bar t}$ spectra and the unfolded CDF spectrum~\cite{Jung:2011zv,Gresham:2011pa,Grinstein:2011dz1}.  
The $\epsilon_i$ are evaluated using the table provided in~\cite{Grinstein:2011dz1}
for the approximate CDF $t\bar t$ pair detection efficiencies, given in bins of $m_{t\bar t} $ and the rapidity difference, $\Delta y =y_t - y_{\bar t }$.
Inclusion of constraints from the CDF $m_{t\bar t} $ spectrum would slow down our $\chi^2$ search procedure substantially, due to evaluation of the $\epsilon_i$.  Fortunately, this turns out to be unnecessary, as the regions of parameter space favored by our current search are consistent  
with the CDF $m_{t\bar t}$ spectra, see examples below.
This is also the case for the predicted LHC $m_{t \bar t}$ spectra.  However, we will see that the LHC is beginning to probe 
the high $m_{t \bar t }$ tail expected in the $Z'$ model.

\begin{table}
\begin{center}
\begin{tabular}{lcc} \hline \hline
scale choice~~ &$\begin{array}{c}~~(m_{Z^\prime},g_{ut})~~\\ \big[\mathrm{Br}(Z'\to u\bar t)=1/4 \big]\end{array}$&$\begin{array}{c}~~(m_{Z^\prime},g_{ut})~~\\ \big[\mathrm{Br}(Z'\to u\bar t)=1/3 \big]\end{array}$\\[4mm] \hline
$\mu=m_t/2$&$(223,0.70)$ & $(225,0.71)$\\ 
$\mu=m_t$&$\begin{array}{c} {\bf p_1}\, \bf(221,0.66)\\ {\bf p_2}\,\bf(323,0.84)\end{array}$ & $(229,0.68)$\\ 
$\mu=2m_t$&$\begin{array}{c}{\bf p_3}\,\bf(222,0.63)\\ {\bf p_4}\,\bf(372,0.91) \end{array}$ &$\begin{array}{c} {\bf p_5}\,{\bf (276,0.74)} \end{array}$\\
$\mu=4m_t$&$\begin{array}{c}(224,0.61)\\(374, 0.88)\end{array}$ &$\begin{array}{c}(286,0.73)\end{array}$\\ \hline \hline
\end{tabular}
\caption{Benchmark points in $(m_{Z^\prime},g_{ut})$     
which minimize the $\chi^2$ for different choices of $\mu$ and ${\rm Br}(Z^\prime \to \bar t u )$, indicated in Figs.~\ref{fig:chi1}--\ref{fig:chi3} with black dots. If two entries are included, the second one corresponds to a higher $m_{Z^\prime}$ benchmark that is marked with a yellow cross in Figs.~\ref{fig:chi2}, \ref{fig:chi3}. Benchmark points $p_1,..,p_5$ (bold) are used in subsequent sections.}
\label{tab:BMP}
\end{center}
\end{table}

In our fits, the NP $t \bar t$ cross sections, and the LO SM cross sections entering the denominators of the $t \bar t$ asymmetries, are evaluated for a wide range of renormalization scales, $\mu = m_t \times (1/2,1,2,4)$, in order to capture the large theoretical uncertainty associated with scale variation at LO.  The corresponding associated $Z'$ production cross sections are evaluated at the scales, $\mu = (m_t + m_{Z^\prime})\times (1/4, 1/2, 1, 2 )$.  The SM cross sections
are listed in the third column of Table~\ref{tab:expsmtheory}. The total SM cross section at the Tevatron, $\sigma^{\rm SM}_{\rm tot}$ (TEV), is identified with the NNLO + NNLL 
evaluation in~\cite{Baernreuther:2012ws} for $m_t = 173.3$ GeV, and 
the total SM cross section at the LHC, $\sigma^{\rm SM}_{\rm tot}$ (LHC), is identified with the approximate NNLO + NNLL evaluation in~\cite{Beneke:2011mq}
for $\sqrt{s} = 7$ TeV and $m_t = 173.3$ GeV.
At this order, the errors associated with scale variation are relatively small.  Thus, in our fits they are absorbed into the overall theoretical uncertainty on the SM cross sections.
The $\Delta \sigma^{\rm SM}$'s entering the $t\bar t$ asymmetries are evaluated at the same renormalization scales as the NP and LO SM cross sections. 
For $\mu = m_t/2 , m_t , 2 m_t$
they are identified with the values given in~\cite{Bernreuther:2012sx} for $m_t = 173.1$ GeV, corresponding to the asymmetries given in Eqs.~(\ref{AFBsm}), (\ref{AFBsmhighlow}) and (\ref{ACsm}).  They are obtained
at NLO including electroweak corrections. In those cases where only an asymmetry is provided in~\cite{Bernreuther:2012sx}, 
we multiply by the corresponding
LO SM cross section to obtain $\Delta \sigma^{\rm SM}$.
For $\mu = 4 m_t $, the $\Delta \sigma^{\rm SM}$'s are given by the sum of their NLO values, obtained using MCFM~\cite{MCNLO},
and estimates of their EW corrections, obtained by extrapolating the ones given in~\cite{Bernreuther:2012sx}.

\begin{figure}
\begin{center}
\includegraphics[scale=0.56]{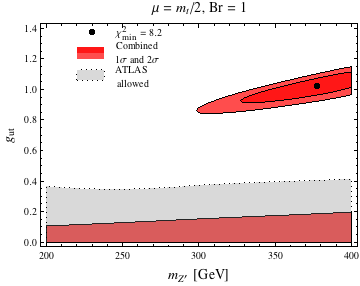}
\includegraphics[scale=0.56]{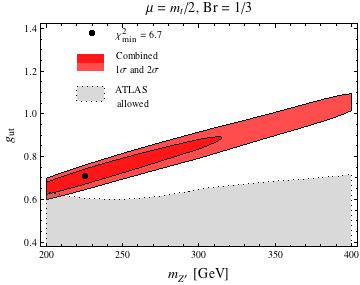}
\includegraphics[scale=0.56]{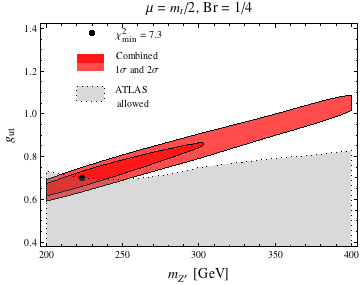}
\caption{$\chi^2$ analysis for $\mu = m_t/2$, $\mathrm{Br}(Z'\to u\bar t) = 1,1/3,1/4$ giving $1\sigma$ and $2\sigma$ preferred regions (red), see text. The gray area represents the region which is not excluded by ATLAS searches for top+jet resonances~\cite{ATLAStj}. The black dot denotes the best fit point. 
}
\label{fig:chi1}
\end{center}
\end{figure}

\begin{figure}
\begin{center}
\includegraphics[scale=0.56]{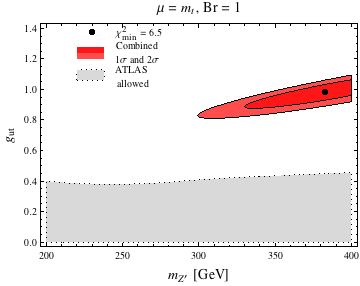}
\includegraphics[scale=0.56]{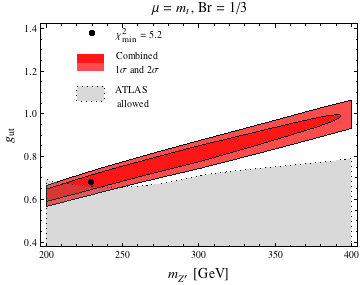}
\includegraphics[scale=0.56]{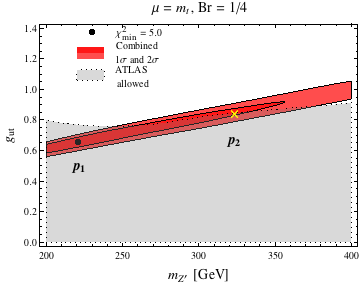}
\caption{Same as Fig.~\ref{fig:chi1} for $\mu = m_t$, with the yellow cross corresponding to the higher $m_{Z^{\prime}}$ benchmark in Table~\ref{tab:BMP}.}
\label{fig:chi2}
\end{center}
\end{figure}

\begin{figure}
\begin{center}
\includegraphics[scale=0.56]{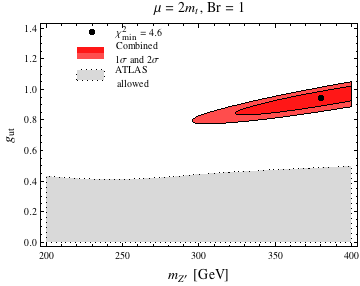}
\includegraphics[scale=0.56]{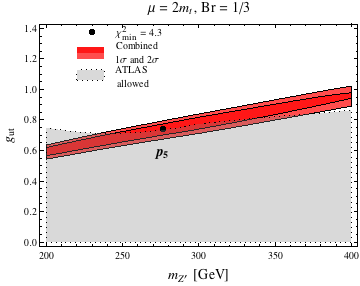}
\includegraphics[scale=0.56]{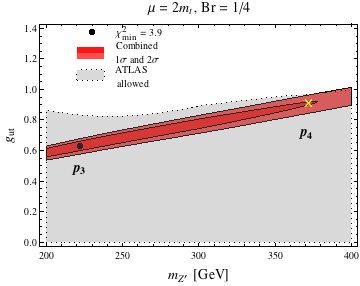}
\caption{Same as Fig.~\ref{fig:chi1} for $\mu = 2 m_t$, with the yellow crosses corresponding to the higher $m_{Z^{\prime}}$ benchmark in Table~\ref{tab:BMP}.}
\label{fig:chi3}
\end{center}
\end{figure}

We exhibit the results of our scans in the $(m_{Z^\prime} , g_{ut} )$ plane in Figs. \ref{fig:chi1}--\ref{fig:chi3}, for three renormalization scales $\mu = m_t /2 , m_t , 2 m_t$. 
For each scale choice, results are shown for three representative branching ratios, ${\rm Br}  (Z^\prime \to \bar t  u)=  1,1/3 ,1/4$.  
For each scan (fixed $\mu$ and ${\rm Br}$), the minimum $\chi^2$ and the corresponding values of $(m_{Z^\prime} , g_{ut} )$ are indicated.
In the plots, the $1\sigma\,(2\sigma)$ region of coverage probability satisfies $\chi^2 - \chi^2_{\rm min} < 2.3\,(5.99)$, corresponding to variation of 2 parameters, $m_{Z^\prime}$ and $g_{ut} $.  
The plots also show the regions which are compatible with the recent ATLAS bounds on top+jet resonance production~\cite{ATLAStj}, to be discussed in the next section.

The best fit points tend to lie at low $m_{Z^\prime} $, near 225 GeV.
Table~\ref{tab:BMP} lists our best fit points, together with a few examples at heavier $m_{Z^\prime} $ near 300 GeV, which are marked with yellow crosses in the plots. The table also includes the best fit points for $\mu = 4 m_t$ and ${\rm Br}=1/4, 1/3$ and a heavier example at this scale for ${\rm Br}=1/4$  (the $\chi^2_{\rm min} = 3.1 $ and 3.3 for ${\rm Br}=1/4$ and 1/3, respectively, and 
$\chi^2 = 5.4$ for the heavier example).
For comparison, for the SM we find the $\chi^2$ values 12.1 ($\mu = m_t/2$), 12.9 ($m_t$), 14.4 ($2 m_t $), 15 ($4 m_t $).   Thus, we find that the $Z'$ model can significantly improve the fit.  In particular, it can lower the $\chi^2$ by up to $\approx 5,8,10,12$
for $\mu = m_t /2, m_t, 2 m_t , 4 m_t$.  

Regardless of scale choice, we find that the $\chi^2$ is minimized for ${\rm Br}  (Z^\prime \to \bar t  u) \approx 1/4$. 
The quality of the fits improves as the renormalization scale is increased, as does the overlap with regions that are consistent with the ATLAS
$tj$ constraint.  
To first approximation, this is because as the scale is increased: {\em i)} the LO NP quantities $\Delta \sigma^{\rm t\bar t,\,Z^\prime}$ and 
$\Delta \sigma^{Z^\prime t}$ decrease as one power of $\alpha_s$, while the LO SM cross sections in the denominators of the $A_{FB}'s$ decrease like $\alpha_s^2$, leading to an overall increase in the forward-backward asymmetries. A byproduct is that at larger $\mu$ the observed asymmetries can be reproduced with smaller couplings;  
{\em ii)} the magnitudes of the LO NP cross sections $ \sigma_{\rm tot}^{t\bar t,\,Z^\prime} $ and $ \sigma_{\rm tot}^{Z^\prime t} $ decrease as one power of $\alpha_s $ which, when combined with the possibility of smaller couplings, makes it easier to satisfy the total cross section bounds.
For example, for ${\rm Br}(Z^\prime \to \bar t  u) \approx 1/4$ and $\mu = m_t$
the minimum $\chi ^2 $ per d.o.f. ($p_1$ in Table~\ref{tab:BMP}) corresponds to a reasonable $p$-value of 17\% or 1.4$\sigma$, whereas for $\mu = 2 m_t$ it improves to 27\% or 1.1 $\sigma$ ($p_3 $ in Table~\ref{tab:BMP}).  Note that we have 3 d.o.f., as we fit 6 observables with 3 parameters, i.e., the coupling, the mass and the branching ratio. 
Reasonable fits are also possible for heavier $Z^\prime$ masses.  For example, 
for $m_{Z^\prime} = 276$ GeV and $\mu = 2 m_t$, ${\rm Br} = 1/3$ ($p_5$ in Table~\ref{tab:BMP}) the
$p$-value is 23\% or 1.2$\sigma$, and the ATLAS $t j $ bound is satisfied.

\begin{figure}
\begin{center}
\includegraphics[scale=0.69]{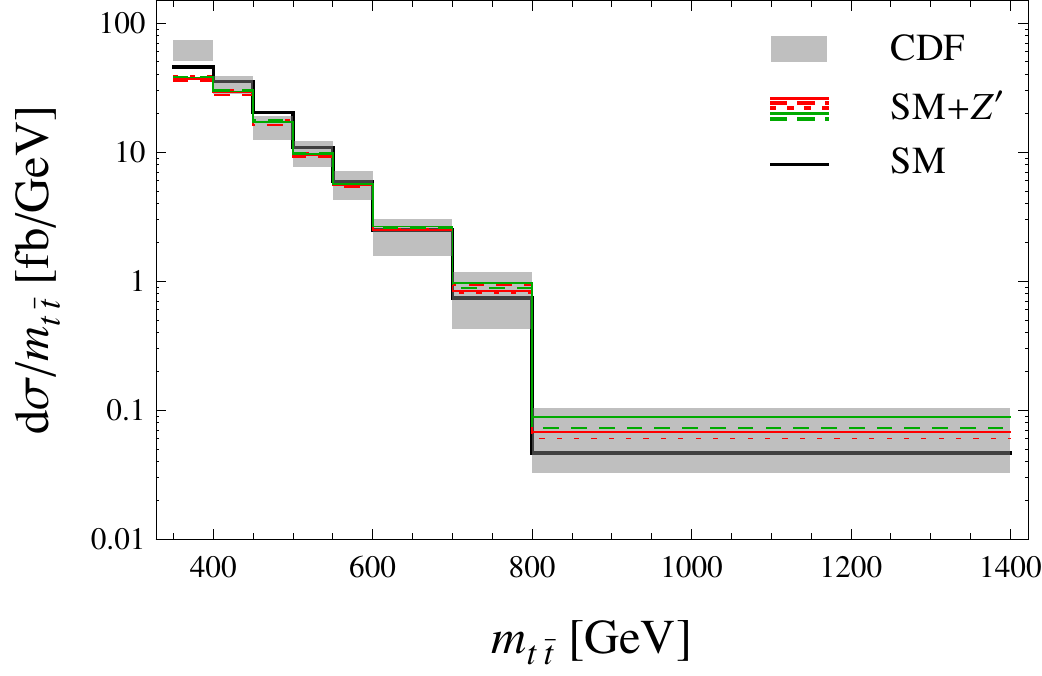}
\includegraphics[scale=0.69]{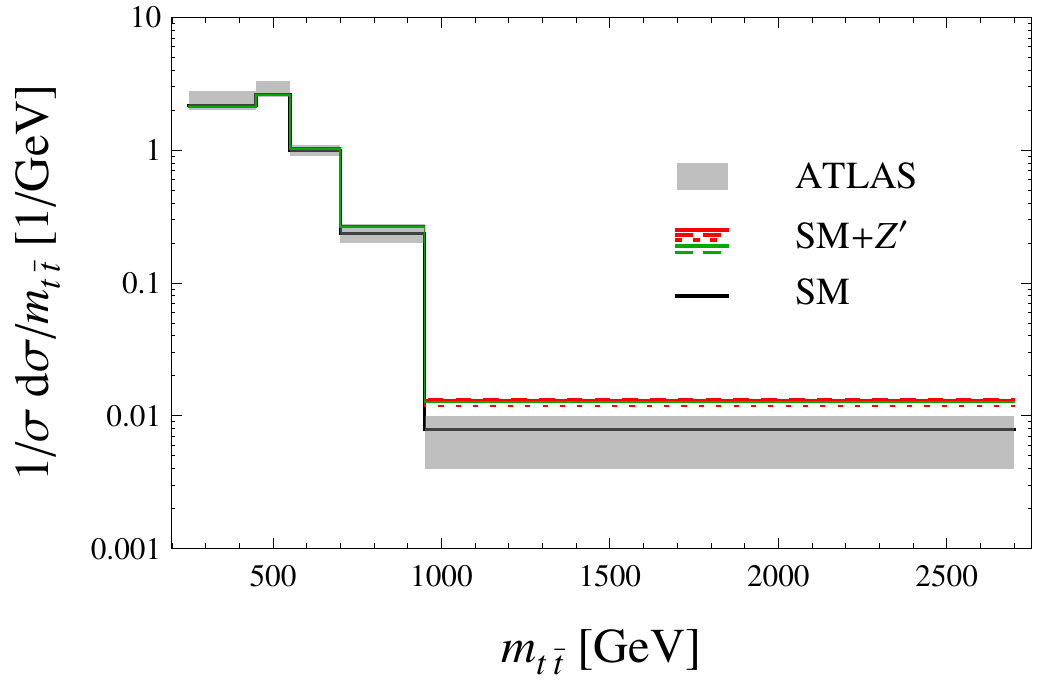}
\caption{The measured Tevatron~\cite{Aaltonen:2009iz} and LHC~\cite{:2012hg} $m_{t\bar t}$ spectra ($1\sigma$ grey bands), the SM (black solid lines), and $p_1 $ (red solid), $p_2$ (red dashed), $p_3 $ (red dotted), $p_4$ (green solid), $p_5$ (green dashed) Table~\ref{tab:BMP} benchmark point predictions.}
\label{fig:spec}
\end{center}
\end{figure}

In Fig.~\ref{fig:spec} we compare the $d\sigma/d m_{t\bar t}$ spectra for the SM and the five benchmarks $p_1 ,.., p_5$ listed in Table \ref{tab:BMP} with the measured Tevatron~\cite{Aaltonen:2009iz} and LHC~\cite{:2012hg} spectra. For the LHC the relative spectrum, $(1/\sigma)\, d\sigma/d m_{t\bar t}$,  is shown. 
The SM predictions are NLO for the Tevatron, 
and 
 NLO+NNLL for the LHC \cite{Ahrens:2010zv}. 
The differences between the SM predictions and the four benchmarks are smaller than the experimental errors (grey bands), with the exception of the last bin in the LHC $m_{t\bar t}$ spectrum.  
The SM exceeds 
the ATLAS collaboration measurement by $\sim 1\sigma$ in this bin, and the five benchmark points exceed it by $\sim 1.5\sigma$ (whether the ``deficit"  relative to the SM is a genuine effect, or is just the result of the last bin typically being statistically challenged
will become clearer once the 8 TeV data is analyzed).
Thus, the LHC is beginning to probe the high mass tail that results from exchange of the $Z'$ in the $t$-channel \cite{AguilarSaavedra:2011vw}.

\begin{figure*}
\begin{center}
\includegraphics[scale=0.50]{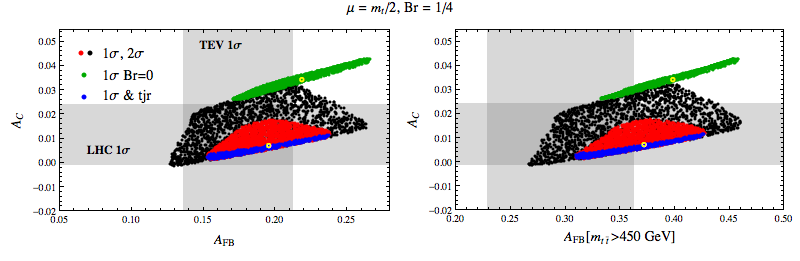}
\includegraphics[scale=0.50]{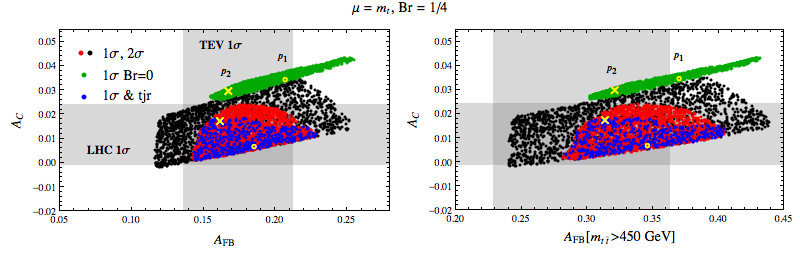}
\includegraphics[scale=0.5]{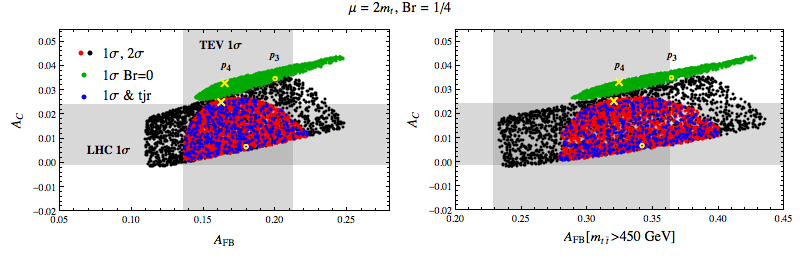}
\includegraphics[scale=0.50]{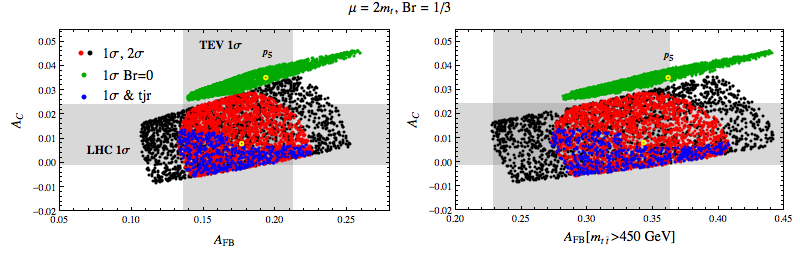}
\caption{ Scatter plots in the  ($A_{FB},A_C$) plane [left] and ($A_{FB}^{\high},A_C$) plane [right], for
variation over $m_{Z^\prime}$, $g_{ut}$ and (i) $\mu=m_t/2$, $\mathrm{Br}= 1/4$, (ii) $\mu=m_t $, $\mathrm{Br}= 1/4$, (iii) $\mu=2 m_t $, $\mathrm{Br}= 1/4$, (iv) $\mu=2 m_t $, $\mathrm{Br}= 1/3$.  Points compatible with the $1 \sigma$ $(2\sigma)$ regions 
in Figs.~\ref{fig:chi1}- 
\ref{fig:chi3} are shown in red (black).  1$\sigma$ region points that are compatible with the ATLAS top+jet resonance bound are shown in blue. The green points are obtained from the $1\sigma$ points 
 by setting $\mathrm{Br}=0$.  A yellow circle corresponds to the $\chi_{\rm min}^2$ point, and its shift upon setting $\mathrm{Br}=0$. The yellow crosses correspond to a second heavier benchmark point, if indicated in Figs.~\ref{fig:chi2}, \ref{fig:chi3}, and its shift for $\mathrm{Br}=0$.  The gray bands are the experimental $1\sigma$ intervals.}
\label{fig:row1}
\end{center}
\end{figure*}

We end this section with our most important result, an illustration in Fig.~\ref{fig:row1}  
of the crucial role played by associated $Z^\prime$ production in reconciling the Tevatron and LHC $t\bar t $ asymmetries.  
The red (black) scatter points in the ($A_{FB},A_C$) and ($A_{FB}^{\high},A_C$) planes correspond to the 1$\sigma$ (2$\sigma$)
regions of coverage probability in the corresponding plots of Figs.~\ref{fig:chi1}--\ref{fig:chi3}.
In addition we show in blue, whenever possible, the subset of points in the $1\sigma$ regions that are also compatible with the ATLAS top+jet resonance bound (see next section for details).
The green points, obtained from the red and blue ones by setting ${\rm Br}(Z^\prime \to \bar t u ) = 0$, clearly illustrate the sizable positive shifts in $A_C $ which occur when associated $Z'$ production is turned off. 
Moreover, while the red/blue $1\sigma$ scatter points tend to lie well inside the experimental $1\sigma$ band for $A_C$, the green points tend to lie above it.

\section{Constraints from top+jet resonance searches, and jet multiplicity distributions}
\label{sec:3}

\subsection{top+jet resonance}
The $Z'$ models we discuss can be searched for by hunting for a bump in the $t+$jet distribution in the $t\bar tj$ final state~\cite{Gresham:2011dg}. Such searches have been performed both at the LHC~\cite{Chatrchyan:2012su,ATLAStj} and the Tevatron~\cite{Aaltonen:2012qn}.  The ATLAS search obtains a bound of $\sigma(pp\to t Z^\prime)+\sigma(pp\to\bar tZ^\prime)\lesssim 23, 14, 7$ pb for $m_{Z^\prime}=200,300,400$ GeV respectively. For a comparison we list in Table \ref{tab:B} the two cross sections, $\sigma(pp\to t Z^\prime)$ and $\sigma(pp\to\bar tZ^\prime)$,  for the four benchmark points. We see that for ${\rm Br}(Z^\prime \to u\bar t)=1/4$ the benchmark points are not excluded. The comparison of the ATLAS bound with the regions in the $g_{ut}$, $m_{Z^\prime}$ parameters space preferred by the $t\bar t$ asymmetry and cross section measurements  are also shown in Figs. \ref{fig:chi1}-\ref{fig:chi3}. The CDF search~\cite{Aaltonen:2012qn} is slightly less sensitive so that for the four benchmarks no bound on ${\rm Br}(Z^\prime \to u\bar t)$ is obtained. The hard cuts in the CMS analysis~\cite{Chatrchyan:2012su} lead to a loss of sensitivity for light $Z^\prime$'s with masses below 400 GeV, making the analysis irrelevant for our case. 

The ``wrong flavor" $Z^\prime\to \bar u t$ decay could also give a contribution to the same-sign top pair  cross section from $pp\to Z' t$ production. By our assumption $Z^\prime$ only decays into $u\bar t$, as is the case in the flavor symmetric models,  so this is not a problem. 
 The experimental bounds from~\cite{Aad:2012bb} are saturated for $Br(Z^\prime \to \bar u t)<3-5\%$, depending on the benchmark (the bounds in~\cite{CDFsame-sign-top} and~\cite{Chatrchyan:2011he} are even less severe). They thus need to be an order of magnitude smaller than the allowed decay channel, $Br(Z^\prime \to u \bar t)\simeq 25\%$.

\begin{table}[t]
\begin{center}
\begin{tabular}{ccccc}\hline\hline
  &\multicolumn{2}{l}{~~$\sigma (p p \to Z^{\prime}t )$ [pb]~~}& \multicolumn{2}{l}{~~$\sigma (p p \to Z^{\prime}\bar t)$ [pb]~~} \\ \hline
        & TEV & LHC &TEV &LHC \\
$p_1$ &0.57& 61.6& 0.57& 5.74\\
$p_2$ &0.19& 40.2& 0.19& 3.13 \\
$p_3$ &0.39& 47.5& 0.39& 4.39\\
$p_4$ &0.079& 25.9&0.079& 1.89\\
$p_5$ &0.24& 40.3& 0.24 & 3.43\\
\hline\hline
\end{tabular}
\caption{The LO $pp\to Z^\prime t$ and $pp\to Z^\prime \bar t$ cross sections at the Tevatron (TEV) and the LHC
for Table~\ref{tab:BMP} benchmarks.
}
\label{tab:B}
\end{center}
\end{table}

\subsection{Jet multiplicities}
Associated $t+Z^\prime \to t\bar t j$ production 
would lead to some deviation in the jet multiplicity distribution for $t\bar t$ events relative to the SM prediction~\cite{Sheltontalk}.
While it would be negligible at the Tevatron (cf. Table \ref{tab:B}), some sensitivity to such deviations might be expected at the LHC.
In Ref.~\cite{CMSjets} such a study was made for $t\bar t$ semileptonic candidate events characterized by the number of $b$-tags, lepton flavor and the number of jets. 
We focus on the double $b$-tagged subsample since it is subject to smaller background contamination.
The subsample is binned in terms of the number of jets.
In the experimental analysis~\cite{CMSjets} a fit to the theoretical prediction was made, allowing the different backgrounds as well as the SM $t\bar t$ contributions to float in their normalization.
Following this procedure, reasonable agreement with the SM Monte Carlo prediction was observed. The statistical error in bins of different jet multiplicities is 
of order a few percent.
However, the background contamination is ${\mathcal O}(30\%)$  or more in the bins with three or fewer jets and is ${\mathcal O}(10\%)$ for the bins with four jets, and five or more jets. 

\begin{figure}[t]
\begin{center}
\includegraphics[scale=0.6]{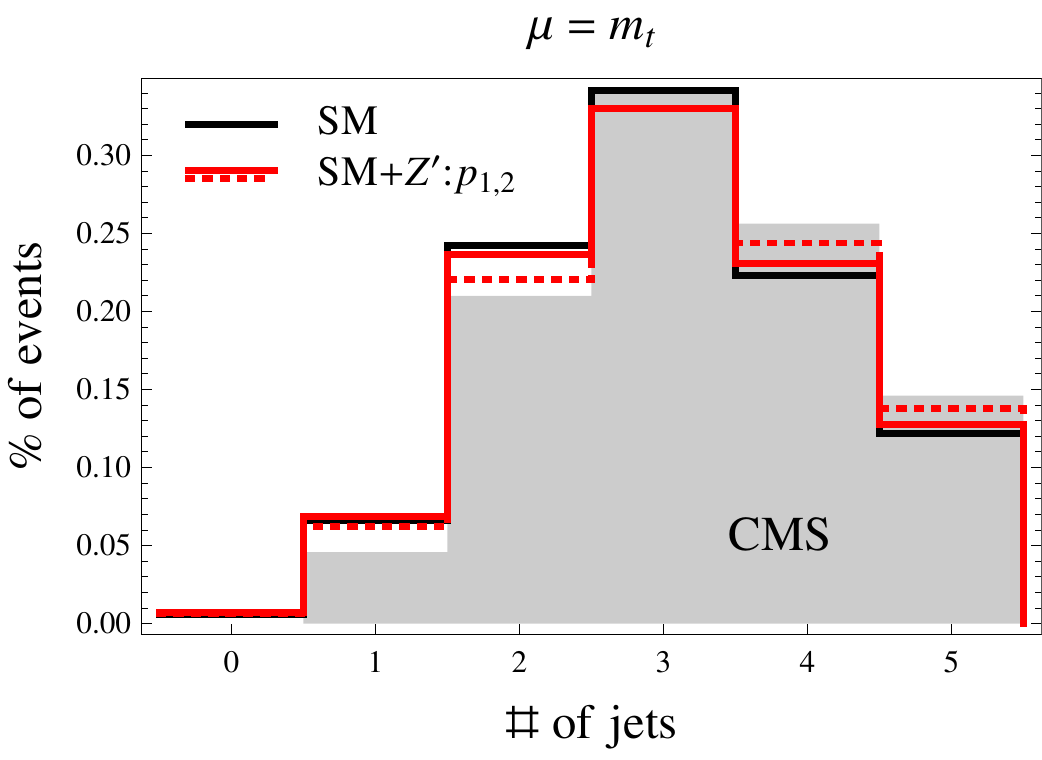}\hspace{1.5cm}
\includegraphics[scale=0.6]{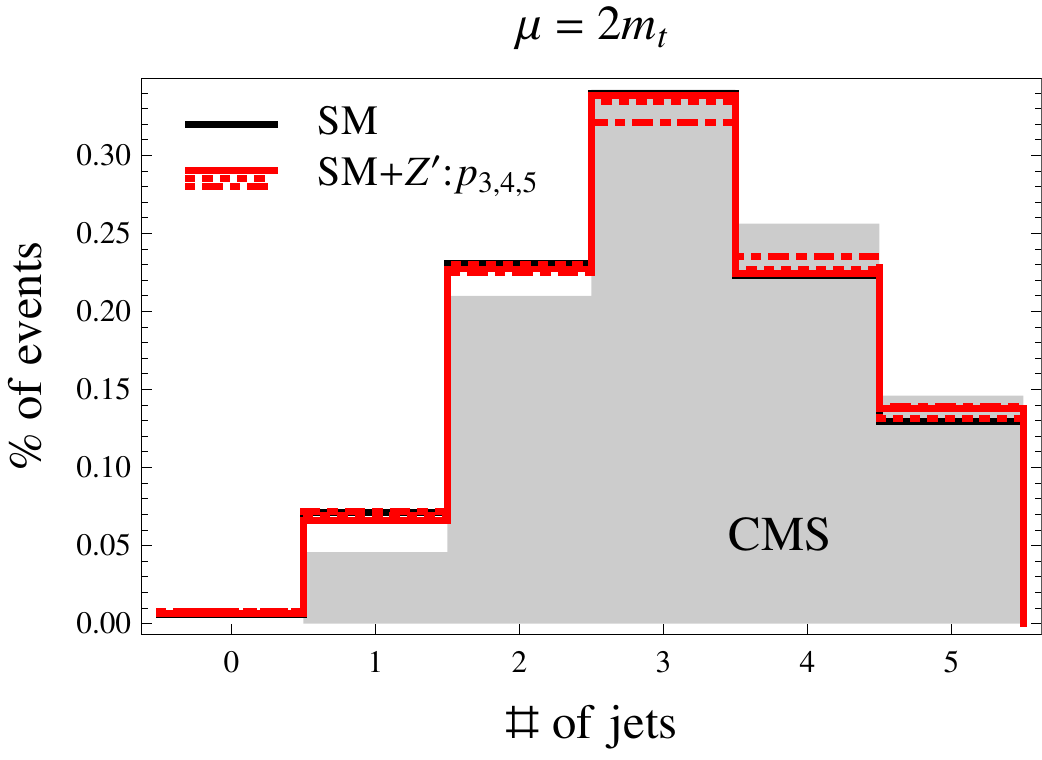}
\caption{Fraction of events with $n=1,...,4$ and $n \ge 5$ jets for $p p \to t \bar t + t \bar t j$ processes with MLM matching for the SM (black) and the five Table \ref{tab:BMP} benchmarks $p_1,..,p_5$ (red). Shaded area represents CMS experimental values obtained from Ref.~\cite{CMSjets}.}
\label{fig:multi}
\end{center}
\end{figure}

In order to check the viability of our mechanism we have performed the following semi-quantitative analysis. Using  {\tt MadGraph5}~\cite{Alwall:2011uj}, {\tt Pythia6.425}~\cite{Sjostrand:2006za} 
and {\tt FastJet}~\cite{Cacciari:2011ma} we have produced an MLM matched $t\bar t+t\bar tj $ sample with and without the new physics contributions, mimicking the jet reconstruction and cuts employed by CMS. The computed jet multiplicities are compared with the experimental values obtained from Ref.~\cite{CMSjets} for  the lepton plus $\mathrm{jets}$ channel (third column of Table 2). The results for the SM and for the five benchmark points in Table~\ref{tab:BMP} are shown in Fig.~\ref{fig:multi}. In none of the bins is the difference between the SM only and the SM + NP samples larger than a few precent.
. 

\begin{table}[t]\begin{center}
\begin{tabular}{lcc}\hline\hline
&~~$pp\to t\bar t + t \bar t j$~~ &~~$p p \to Z^\prime t (\bar t) \to t \bar t X$~~
 \\ \hline
SM & 0.12 & -- \\ \hline
$\mathrm{SM} + Z^{\prime} (p_1)$~~ & 0.13 & 0.22 \\
$\mathrm{SM} + Z^{\prime} (p_2)$ & 0.14 & 0.31 \\
$\mathrm{SM} + Z^{\prime} (p_3)$ & 0.13& 0.25 \\
$\mathrm{SM} + Z^{\prime} (p_4)$ & 0.14  & 0.36  \\ 
$\mathrm{SM} + Z^{\prime} (p_5)$ & 0.14  & 0.31 \\ \hline\hline
\end{tabular}\end{center}
\caption{Fraction of MLM matched $t\bar t+t\bar tj$ events with number of jets equal to 5 or more for the SM and $\mathrm{SM}+Z^\prime$ for the Table~\ref{tab:BMP} benchmark points $p_1,..,p_5$.}
\label{tab:multi}
\end{table}

Of special interest for $t+Z^\prime\to t\bar t j$ searches are the bins with large jet multiplicities. We thus analyze in more detail the fraction of events with 5 or more jets, with the results collected in Table~\ref{tab:multi}. The fraction of events that have more than 5 jets is 0.12 for the SM, and is between 0.13 and 0.14 for the five Table~\ref{tab:BMP} benchmarks $p_1,..,p_5$, well within the experimental errors. Isolating only the events with $Z^\prime$ associated production, the fraction is significantly larger, between $0.22$ and $0.36$ depending on the benchmark point. Because the $pp\to Z^\prime t \to t\bar t X$ cross sections are substantially smaller than the total $t \bar t$ cross section, one sees that the impact on the combined SM+NP jet multiplicity is small.

\section{Contraints from Atomic Parity Violation}
\label{sec:4}

New massive bosons with parity violating couplings to light quarks can modify at one-loop the nuclear weak charge ($Q_W$) measured in atomic parity violation (APV) experiments. Most constraining are the precise measurements with cesium (${}^{133}$Cs)~\cite{Wood:1997zq}, which are in reasonable agreement with the most recent SM predictions~\cite{Dzuba:2012kx}
\beq
Q^{\rm exp}_W({\rm Cs}) - Q^{\rm SM}_W({\rm Cs}) = 0.65(43)\,.
\label{eq:QWexp}
\eeq
This result can be interpreted in terms of constraints on the effective couplings of the SM $Z$ boson to the light $u$ and $d$ quarks.
It was argued in~\cite{KZAPV} that this leads to severe constraints on the $Z'$ explanation of the $A_{FB}$ puzzle. The argument is based on a calculation performed in a toy $SU(2)_X$ model~\cite{Jung:2009jz}, which contains two new spin-1 states coupling to right-handed up-type quarks: $Z'$ ($V'$ in the notation of~\cite{KZAPV}) with mostly flavor violating ($\bar u - t$) couplings, and $Z''$ ($Z'$ in the notation of~\cite{KZAPV}) with mostly flavor diagonal ($\bar u - u$ and $\bar t - t$) couplings. In this setup only the $Z'$ is directly relevant for $A_{FB}$. On the other hand, the strong APV constraints on the model are dominated by the $Z''$ contributions. 

\begin{figure}
\begin{center}
\includegraphics[scale=0.55]{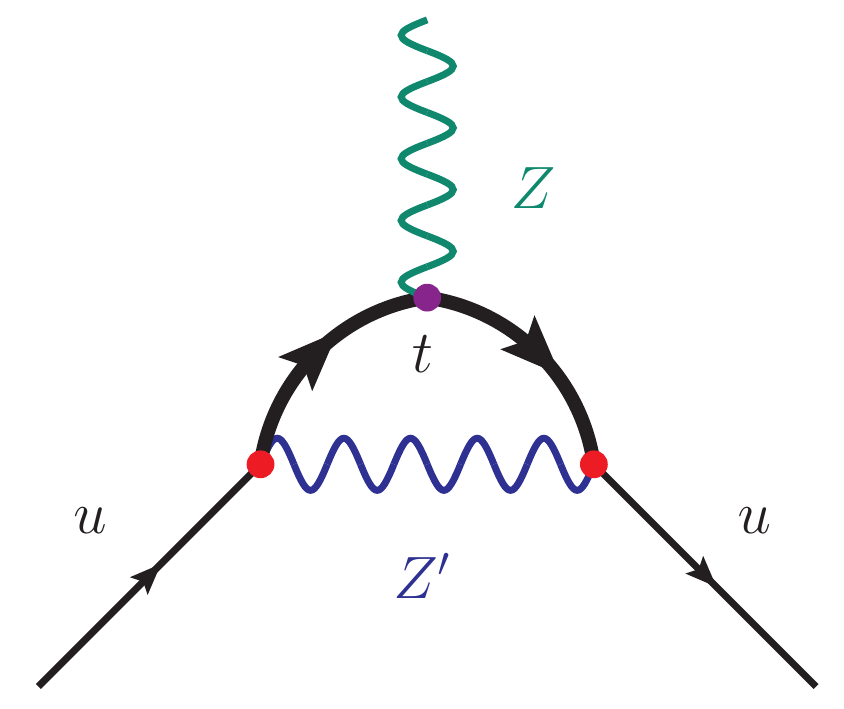}
\caption{\label{fig:APV_feyn}
Feynman diagram for the $Z'$ contribution to the weak nuclear charge via the modification of the effective $\bar u_R Z  u_R$ coupling.
 }
\end{center}
\end{figure}

\begin{figure}
\begin{center}
\includegraphics[scale=1.2]{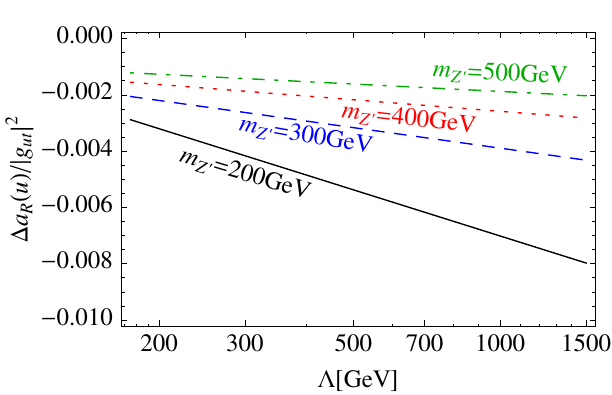}
\caption{\label{fig:APV_EFT}
The $Z'$ induced $\Delta a_R (u)$ contributing to APV in the minimal effective non-renormalizable model, as a function of the effective theory cut-off scale $\Lambda$ for several representative $Z'$ masses.   
 }
\end{center}
\end{figure}

To demonstrate this we evaluate the modification of $Q_W$ in the minimal model~\eqref{eq:lagrangian}. This can be viewed as the limit of the $SU(2)_X$ model with $Z''$ heavy enough that it decouples (another possibility is to enlarge the symmetry group to include a $U(1)$ factor, see below). We denote by $\Delta a_R (u)$ the shift 
in the effective $\bar u_R Z  u_R$ coupling
induced by the $Z'$ loop, cf. Fig.~\ref{fig:APV_feyn}. Since the setup is non-renormalizable we impose a hard cut-off scale, $\Lambda$. This very roughly corresponds to the mass scale of the new degrees of freedom that will regulate the loop divergences found in the effective theory. In this way we obtain~\cite{KZAPV}
\beq
\begin{split}
\Delta a_R(u) &= \frac{-|g_{ut}|^2}{16\pi^2} \frac{m_t^2}{m_{Z'}^2} \left[ F(m_t^2/m_{Z'}^2) + \tfrac{1}{4} \log\left( \Lambda^2 / m_t^2 \right)  \right]\,,
\label{eq:aRuEFT}
\end{split}
\eeq
where $F(x) \equiv (x-1-\log x)/(1-x)^2$. The resulting shift to the nuclear weak charge is given by
\beq
\Delta Q_W = -2 (2 Z_n + N_n) \Delta a_R(u),
\eeq
where $Z_p$ and $N_n$ are the number of protons and neutrons in a nucleus, respectively.
One can thus translate Eq.~\eqref{eq:QWexp} to a 1-$\sigma$ preferred region for $\Delta a_R(u) = -0.17(11)\times 10^{-2}$\,. This is to be compared with the predictions in the minimal model  as shown in Fig.~\ref{fig:APV_EFT}.  We have factored out the trivial dependence  on the coupling  $g_{ut}$ (note that in our best-fit regions $|g_{ut}|\lesssim 1$) and plot $\Delta a_R(u)/|g_{ut}|^2$ as a function of the effective theory cut-off $\Lambda$ for several representative $Z'$  masses. We see that for $m_{Z'}\gtrsim 300$\,GeV the APV bound can be satisfied even for cut-off scales above 1~TeV. 

\begin{figure}[t]
\begin{center}
\includegraphics[scale=1.2]{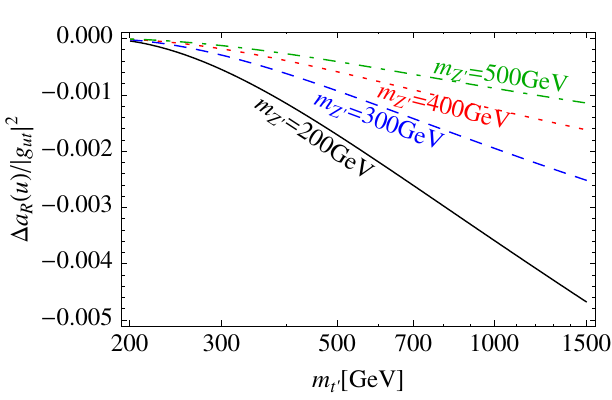}
\caption{\label{fig:APV_tp}
The $Z'$ induced $\Delta a_R (u)$ contributing to APV in the renormalizable model, as a function of the vector-like quark mass $m_{t'}$ for several representative $Z'$ masses (see text for details).    
 }
\end{center}
\end{figure}

The conclusions drawn in the effective non-renormalizable model can be further verified in a renormalizable ultraviolet completion of the $SU(2)_X$ model~\cite{KZAPV}, where the $Z'$ (and $Z''$) loop divergences are cut-off by contributions of an additional heavy vector-like quark ($t'$) with the SM gauge quantum numbers of the right-handed top. This gives 
\beq
\Delta a_R(u) = \frac{-|g_{ut}|^2}{16\pi^2} \frac{m_t^2}{m_{Z'}^2} F_1(m_t^2/m_{Z'}^2,m_{t'}^2/m_{Z'}^2) +\ldots   \,.
\eeq
The dots denote additional $Z''$ contributions, while
\begin{align}
F_1(x,y) \equiv &- \frac{1}{4} \Bigg[ 2 + \frac{6-3x-3y}{(1-x)(1-y)}  \notag \\ 
& + \frac{(x^2-2x + 4)\log x}{(1-x)^2} + \frac{(2x^2-8x)\log x}{(1-x)(x-y)} \notag \\
&+ \frac{(y^2-2y + 4)\log y}{(1-y)^2} + \frac{(2y^2-8y)\log y}{(1-y)(y-x)} \Bigg]\,.
\end{align}
In the $m_{t'}\gg m_{Z'}$ limit the above expression reproduces the $\log \Lambda$ dependence in Eq.~\eqref{eq:aRuEFT} with $\Lambda\equiv m_{t'}$. Again considering this as an illustrative setup (equivalent to the full $SU(2)_X$ model with a decoupling flavor diagonal $Z''$) we plot $\Delta a_R(u)/|g_{ut}|^2$ as a function of the $t'$ mass ($m_{t'}$) for several representative $Z'$  masses in Fig.~\ref{fig:APV_tp}. 
We observe that the $Z'$ contributions in the renormalizable model are even smaller than estimated in the effective theory and that the current APV constraints can be perfectly satisfied even for $m_{Z'}\simeq 200$\,GeV (where the $A_{FB}$ fit prefers $|g_{ut}|\sim 0.6$) and $t'$ masses well above 1~TeV.

Finally, let us briefly discuss the viability of decoupling $Z''$ contributions to $\Delta a_R(u)$ in the $SU(2)_X$ model used above. The first approach to obtain this limit is to make the $Z''$ heavy. However, the required hierarchical $SU(2)_X$ breaking pattern requires large scalar representations~\cite{Jung:2011zv}. 
Another possibility is to impose an OZI-like alignment of flavor diagonal vector mass eigenstates such that they  couple predominantly to the $\bar u-u$ or $\bar t-t$ current and not both simultaneously, thus suppressing the flavor diagonal contributions to $\Delta a_R(u)$. For example, one can enlarge the symmetry group to include extra $U(1)$ factors (i.e. to $SU(2)_X\times U(1)_X$ or $U(2)_X$). An explicit realization of the OZI alignment is then possible already with two scalar doublets breaking the symmetry group completely, by judiciously  choosing their $U(1)_X$ charges. The desired OZI alignment would also follow 
dynamically from ``$\omega-\phi$" mixing in a strong interaction realization of the $Z'$ model \cite{stronginteractions} in which, as previously noted, the $Z'$ is the composite analog of 
a $K^*$ meson residing in a flavor nonet.  The mixing of the $[\bar t t]$ flavored ``$\phi$" and 
the $[u\bar u + \bar c c]$ flavored ``$\omega$" 
would be suppressed, as in QCD.

\section{Lepton based asymmetries}
\label{Lepton}
An interesting Tevatron observable that is sensitive to the chiral nature of the NP model is the lepton based asymmetry
\beq
A_l\equiv \frac{\sigma_{l^+}(\eta_{l^+}>0)-\sigma_{l^-}(\eta_{l^-}>0)}{\sigma_{l^+}(\eta_{l^+}>0)+\sigma_{l^-}(\eta_{l^-}>0)},
\eeq 
where $\sigma_{l^\pm}(\eta_{l^\pm}>0)$ are the $t\bar t$ cross sections for semileptonic $t$ decays in which the $l^\pm $ lepton pseudorapidity is positive. Similarly, the di-lepton asymmetries can be defined. The inclusive asymmetry has been measured to be $A_l=(15.2\pm4.0)\%$ by the D0 collaboration \cite{AFBD0}, and $A_l=(6.6\pm2.5)\%$ by the CDF collaboration \cite{AFBCDF1}. The CDF collaboration has also measured the asymmetry in two $m_{t\bar t}$
bins, obtaining $A_l(m_{t\bar t}< 450 {\rm~GeV})=(3.7\pm3.1)\%$ and $A_l(m_{t\bar t}> 450 {\rm~GeV})=(11.6\pm4.2)\%$. The SM predictions are $A_l^{SM}=(3.6\pm0.2)\%$, $A_l^{SM}(m_{t\bar t}>450{\rm ~GeV})=(6.4\pm0.5)\%$ and $A_l^{SM}(m_{t\bar t}<450{\rm~GeV})=(1.9\pm0.1)\%$ \cite{Bernreuther:2012sx}.

The lepton based asymmetry is an important observable because it gives an independent probe of the NP model explanations for the $A_{FB}$ measurements. In order to understand the physics that determines $A_l$ in a qualitative manner it is best to consider two kinematical regions for 
the $t\bar t$ invariant mass: near threshold, and well above threshold. Near threshold $A_l$ is predominantly controlled by the chirality of the initial state quarks \cite{Falkowski:2011zr}. Well above the threshold it is determined by the polarizations of the tops \cite{Agashe:2006hk,Godbole:2010kr,Krohn:2011tw}.  In our framework the $Z'$ couples to right-handed quarks.  Thus, we expect a negative value for $A_l$ near threshold, from the interference between the SM amplitude 
and the NP $t$-channel 
amplitude. The contribution to $A_l$ from associated $Z'$ production is small due to its small cross section at the Tevatron. 
At high $m_{t\bar t}$ we expect a positive value of $A_l$ because more right-handed 
tops are produced from the NP interactions. This implies that there will be some cancelation between these two kinematical regions in the inclusive $A_l$. Therefore, the exact value of the inclusive asymmetry will be relatively sensitive to the details of the NP parameter space. 

Numerically, employing the prescription in (\ref{AFB}),
we find that the inclusive values of $A_l$ lie between $4\%$ to $6\%$ for all of our benchmark points ($p_1 ,..,p_5$),
consistent with the CDF measurement and somewhat lower than the D0 measurement.
We also find that $A_l (m_{t\bar t}< 450 {\rm~GeV})\sim -7$ to $-8\%$ and $A_l(m_{t\bar t}>450 {\rm~GeV})\sim 20-23\% $ for all of our benchmark points ($p_1 ,..,p_5$).  Thus, the tendency of $A_l$ to increase with $m_{t\bar t}$ is indeed present in the $Z'$ models, but the rise is more pronounced than observed by CDF. 

\section{Discussion}
\label{sec:5}

In conclusion, we have shown that associated production of a vector state, $pp\to t V$, where $V\to \bar t j$, gives a negative contribution to the $t\bar t$ charge asymmetry $A_C$ at the LHC, while it has a negligible effect on the forward-backward asymmetry $A_{FB}$ measured at the Tevatron.  As is well known, the same underlying interaction also gives rise to positive contributions to $A_{FB}$ and $A_C$,  via exchange of the vector in the $t$-channel.  Thus, $A_C$ can have SM like values, while 
$A_{FB}$ can be anomalously large and positive, in agreement with experiment.
We have discussed this quantitatively for the example of a $Z^\prime$, where the $Z^\prime$ decays to $u\bar t$ roughly $25\%$ of the time.  The main result is summarized in Fig.~\ref{fig:row1}, showing a substantial negative shift in $A_C$ and a much smaller shift in $A_{FB}$, bringing the $Z^\prime$ models into agreement with their measured values. 

Finally, we have shown that the $Z'$ model is not ruled out by other searches: the top+jet resonance search in the $t\bar tj$ final state, the jet multiplicity distributions in $t\bar t$ production, and atomic parity violation measurements. It would be interesting to search for the 
dominant $Z^\prime$ decay channel candidates, e.g., invisible final states, or $t+ n$ jet final states. A test of the $t-$channel $Z'$ exchanges is provided by lepton based asymmetries.

{\bf Note added:} While our work was being submitted to the arXiv, similar conclusions appeared in Ref. \cite{'Alvarez:2012ca}.

\begin{acknowledgments}
We thank Yonit Hochberg and Jessie Shelton for discussions and V. Ahrens for communication regarding the SM predictions for the  $t\bar t$ cross sections. J.D. thanks C. Duhr and O. Mattelaer for enlightening discussions as well as the Lawrence Berkeley National Laboratory and University of Cincinnati, where part of this work was completed, for their hospitality. 
We thank the organizers of the NPKI Workshop ``Top physics and electroweak symmetry breaking in the LHC era", Seoul, February 2012, for their hospitality in the initial stage of this work. The work of J.D. and J.F.K. was supported in part  by the Slovenian Research Agency.  A.L.K was supported in part by DOE grant FG02-84- ER40153.
G.P. is the Shlomo and Michla Tomarin development chair, supported by the grants
from Gruber foundation, IRG, ISF and Minerva.
JZ was supported in part by the U.S. National Science Foundation under CAREER Award PHY-1151392.
\end{acknowledgments}

\end{document}